\documentclass[runningheads]{llncs}
\usepackage[T1]{fontenc}
\usepackage{graphicx}
\usepackage{soul}
\usepackage{amsmath,amssymb}
\usepackage{tabularx}
\usepackage[dvipsnames]{xcolor}
\usepackage{graphicx}
\usepackage{lipsum}
\usepackage{amsmath,amssymb}
\usepackage{caption}
\usepackage{subcaption}
\usepackage{tabularray}
\usepackage{algorithm,algorithmic}

\begin{document}
\title{COVID-CT-H-UNet: a novel COVID-19 CT segmentation network based on attention mechanism and Bi-category Hybrid loss}
%
%
\author{Anay Panja\inst{1} \and
Somenath Kuiry\inst{2}\orcidID{0000-0002-8462-5547} \and
Alaka Das\inst{2}\orcidID{0000-0003-2023-6566} \and Mita Nasipuri\inst{3}\orcidID{0000-0002-3906-5309} \and Nibaran Das\inst{3}\orcidID{0000-0002-2426-9915}}
\authorrunning{Panja et al.}
%
\institute{RCC Institute of Information Technology, Kolkata, India  \\
\and
Department of Mathematics, Jadavpur University, Kolkata-32, West Bengal, India \and Department of CSE, Jadavpur University, Kolkata-32, West Bengal, India 
\email{anaypanja30@gmail.com }\\ 
\email{mitanasipuri@gmail.com}\\
\email{\{skuiry.math.rs, alaka.das, nibaran.das\}@jadavpuruniversity.in}}
\maketitle              
\begin{abstract}
Since 2019, the global COVID-19 outbreak has emerged as a crucial focus in healthcare research. Although RT-PCR stands as the primary method for COVID-19 detection, its extended detection time poses a significant challenge. Consequently, supplementing RT-PCR with the pathological study of COVID-19 through CT imaging has become imperative. The current segmentation approach based on TVLoss enhances the connectivity of afflicted areas. Nevertheless, it tends to misclassify normal pixels between certain adjacent diseased regions as diseased pixels. The typical Binary cross entropy(BCE) based U-shaped network only concentrates on the entire CT images without emphasizing on the affected regions, which results in hazy borders and low contrast in the projected output. In addition, the fraction of infected pixels in CT images is much less, which makes it a challenge for segmentation models to make accurate predictions. In this paper, we propose COVID-CT-H-UNet, a COVID-19 CT segmentation network to solve these problems. To recognize the unaffected pixels between neighbouring diseased regions, extra visual layer information is captured by combining the attention module on the skip connections with the proposed composite function Bi-category Hybrid Loss. The issue of hazy boundaries and poor contrast brought on by the BCE Loss in conventional techniques is resolved by utilizing the composite function Bi-category Hybrid Loss that concentrates on the pixels in the diseased area. The experiment shows when compared to the previous COVID-19 segmentation networks, the proposed COVID-CT-H-UNet's segmentation impact has greatly improved, and it may be used to identify and study clinical COVID-19.
\end{abstract}
\section{Introduction} 
The COVID-19 outbreak struck the globe in December 2019. Instances of COVID-19 have been discovered so far in $197$ different nations \cite{miller20202019}. Around $560$ million cases were confirmed and approximately $6$ million patients died \cite{paules2020coronavirus}. Thus it presents a significant obstacle to global health care. Since COVID-19 currently has no viable treatments, early detection of COVID-19 becomes crucial. The industry acknowledges RT-PCR (Reverse Transcription and Polymerase Chain Reaction) detection as the foremost standard for identifying COVID-19. Nevertheless, amid the COVID-19 outbreak, there is frequently a shortage of suitable detection equipment, and the detection process may take up to 6 hours. In order to address these challenges, computed tomography (CT) and X-ray imaging technologies are implemented as complementary methods to RT-PCR testing.

CT diagnostic technology is a prevalent method for diagnosing lung diseases \cite{sluimer2006computer}. In comparison to X-ray imaging, computed tomography (CT) imaging is more adept at clearly depicting features such as ground glass shadows, pulmonary nodules, pleural effusion, and other anomalies in lung lesions. It proves particularly suitable for medical image segmentation. The segmentation of specific areas from CT images plays a crucial role in aiding the diagnosis of COVID-19. The delineated features can be leveraged to quantify the severity of COVID-19 and calculate the infected area. However, during the COVID-19 outbreak, there was often a shortage of doctors available to annotate CT scans due to the high number of patients. Consequently, the incorporation of deep learning technology in COVID-19 CT image segmentation becomes imperative to enhance the efficiency of the process.

Deep learning has become increasingly popular in recent years for segmenting the major COVID-19 sectors. Examples of such works are, FGCNet proposed by Wang et al. \cite{wang2021covid}, Inf-Net proposed by Fan et al. \cite{fan2020inf}, TV-UNet proposed by Shervin Minaee et al. \cite{saeedizadeh2021covid}, and SCTV-UNet proposed by Xiangbin Liu et al. \cite{liu2023sctv}. These models utilize a conventional encoder-decoder-based (U-shaped) network with Binary cross entropy (BCE)loss. Although the segmentation performance of COVID-19 CT images has been significantly enhanced by these above-mentioned networks, there are some drawbacks to be noted. Firstly, the BCE loss function with the conventional U network creates the issue of hazy borders for some CT images having larger backgrounds than the foreground as pointed out in \cite{liu2023sctv}. Also, there are several anomalies in the CT scans of the COVID-19 CT segmentation dataset such as CT scans with a backdrop that is more substantial than the foreground. In this paper, we propose the COVID-CT-H-UNet network architecture which is based on the similar encoder-decoder-based popular segmentation network U-Net \cite{ronneberger2015u} with attention and a novel hybrid loss function, to overcome the aforementioned issues. We have incorporated the attention module at the decoder layer to acquire more visual layer information from the latent space, which will help to solve the issue of mistaken identification between nearby lesion sites. Also, the boundary ambiguity issue is resolved with the proposed Bi-category Hybrid loss(Bi-H loss) function which is a hybrid combination of two popular categories of the loss function. In summary, the contributions of this paper are as follows:
\begin{enumerate}
    \item This paper proposes a novel COVID-19 segmentation network COVID-CT-H-UNet paired with the skip connections attention mechanism at the decoder layer. 
    \item The novel composite loss function Bi-category Hybrid Loss is proposed in this work, which is a hybrid of two categories of the loss function. 
    \item The approach proposed in this study performs more effective segmentation both qualitatively and quantitatively on the COVID-19 CT segmentation dataset when compared to current methods.
\end{enumerate}

The related work of this article is discussed in the next section of this paper. The approaches applied in this paper are explained briefly in the third section. Also, the network architecture, the dataset description, and the attention mechanism are discussed in the third section. In Section IV, we discussed and analyzed the overall results along with any important experimental findings. Finally, we gave the conclusion in the last section.

\section{Related work}
The segmentation network of medical images, the attention mechanism, and the CT segmentation network of COVID-19 are discussed in this section.

\subsection{The Segmentation Networks for Medical Images}
The subject of medical image segmentation has advanced since standard segmentation networks like FCN  \cite{long2015fully}, U-Net  \cite{ronneberger2015u}, U-Net++ \cite{zhou2018unet++}, and Deeplap  \cite{chen2018encoder} were proposed. 
There are numerous improvements of U-Nets also developed. For example, The MultiResUNet network  \cite{ibtehaz2020multiresunet} was created to address the U-Net \cite{ronneberger2015u} network's flaws, Li et al.'s \cite{li2018h} proposed H-DenseUNet  \cite{li2018h}, a segmentation network for the liver and hepatic cancers where a  3D network receives the concatenated 2D DenseUNet  \cite{li2018h} prediction outputs and the original 3D input to extract inter-slice characteristics.
Some networks decide to enhance the loss function while segmenting medical images. The Lov'asz Softmax loss function \cite{berman2018lovasz} was developed by Maxim Berman et al. and is commonly utilized in medical image segmentation because of its superior performance. Additionally, the weighting loss function of Inf-Net \cite{fan2020inf}, the DTV loss function of SCTV-UNet \cite{liu2023sctv}, and the TV loss function of TV-UNet \cite{saeedizadeh2021covid} have proved successful in the segmentation of COVID-19 CT images. As a result, one of the primary research for medical picture segmentation is the development of loss function.

\subsection{Attention Mechanism with convolutional Neural Networks}
Attention mechanisms are becoming more prominent in convolutional neural networks. Hu et al. presented the SENet  \cite{hu2018squeeze}: squeeze excitation network, which acquired the correlation between channels through squeeze and excitation operations and filtered the attention for channels. In contrast to SENet, which focuses solely on channel attention, Lee et al. introduced CBAM  \cite{woo2018cbam}: Convolution Block Attention Module. This work uses the attention mechanism to focus on relevant characteristics, suppress unimportant ones, and boost the expressiveness of features. CBAM  \cite{woo2018cbam} modules are commonly utilized in CNN network architectures because they can be introduced into most CNN structures with insignificant parameters. Attention Networks are devoted to eliminating the problem of duplicate use of computer parameters and resources that is caused by repeated extraction of comparable low-level features from models in cascade, according to Jo Schlemper et al. However, the traditional convolutional neural network has several drawbacks when it comes to target structures with considerable changes in texture, shape, and size between subjects. Chen et al.  \cite{chen2021transunet} introduced TransUNet, which integrates natural language processing Transform  \cite{wang2019learning} with traditional U-Net \cite{ronneberger2015u}. The encoder is the Transform component, and the decoder is the U-Net \cite{ronneberger2015u} part. By recovering local spatial information, a transform decoder can gain additional details. Similarly, Transform modules replace skip connections and channel stacks to improve the representation of coding characteristics.

\subsection{COVID-19 segmentation network}

Amidst the global COVID-19 pandemic, the utilization of COVID-19 CT segmentation networks for detecting the virus has gained increasing prominence. Notably, substantial progress has been made in the development of COVID-19 CT segmentation networks over the past two years. Fan et al. \cite{fan2020inf} introduced the Inf-Net network architecture in 2020, incorporating a parallel partial encoder, reverse attention, and an edge detection module. The edge detection module addresses challenges related to the identification of the ground glass shadow area boundaries, which are often difficult due to low contrast and a fuzzy appearance. To tackle the issue of missing data in the COVID-19 dataset, a semi-supervised network, Semi-Inf-Net \cite{fan2020inf}, was proposed as an extension of Inf-Net.

The adoption of Inf-Net and Semi-Inf-Net has significantly enhanced metrics like Dice, Sensitivity, and Specificity in COVID-19 CT segmentation, particularly within the COVID-SemiSeg dataset. Subsequently, Lu et al. \cite{lu2022multiscale} presented a multiscale codec network based on CT image segmentation for human lung diseases, specifically those related to COVID-19, building upon the foundation of Inf-Net. This approach leverages multi-scale decoded information to enhance the accuracy of COVID-19 lesion segmentation.

\section{Methodology}
\subsection{Network structure}
\begin{figure}
\centering
\includegraphics[width = \textwidth]{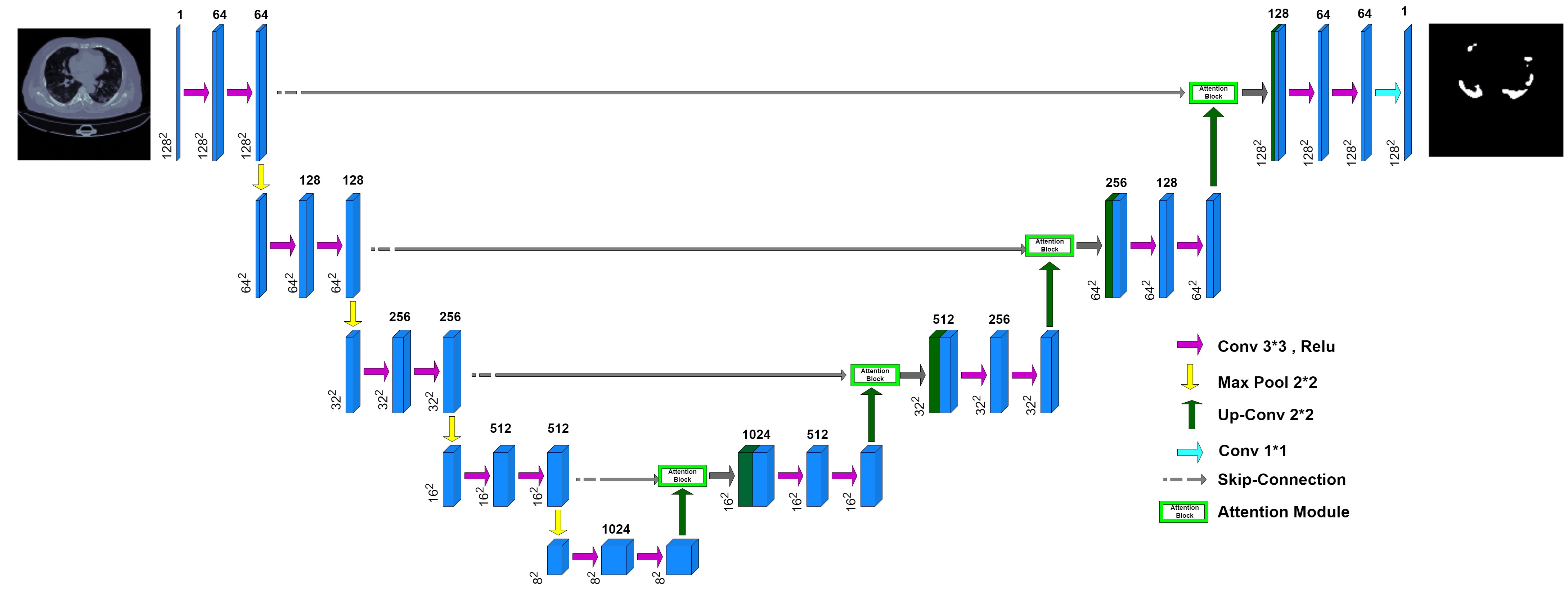}
\caption{The UNet architecture}
\label{unet}
\end{figure}

This paper proposes the COVID-CT-H-UNet network architecture, which is inspired by SCTV-UNet \cite{liu2023sctv} and TV-UNet \cite{saeedizadeh2021covid}. The framework’s main body uses the standard encoder-decoder-based segmentation model U-Net \cite{ronneberger2015u}. The feature extraction module (encoder part) is responsible for extracting relevant feature information from the COVID-19 CT images through repeated Convolution and Max-Pool operations. All the extracted feature information is mapped into the latent space which is the bottom part of the model. From this latent space, the feature fusion module (decoder part) tries to extract extensive information by repeating deconvolution, feature splicing, and two-layer convolution procedures to create the segmented image. One of the key characteristics of U-Net is the unique skip connections between the encoder layer and decoder layers. These skip connections can transmit useful information from As shown the appropriate low-resolution layer of the encoder to the decoder, allowing the network to capture high-resolution features. The network architecture is shown in Fig \ref{unet}, 

In this paper, an attention mechanism is added to the network’s skip connection. Here, four attention modules are alternatively added to the four skip connections. By making features more relevant, it enables the network to prioritize informative areas while reducing noise. Accuracy, resistance to variability, and adaptation to various object sizes and shapes are all improved as a result. Additionally, the attention mechanism aids in the integration of global and local contexts and helps to identify distant relationships for coherent and contextually aware segmentation.
The next section illustrates this attention mechanism. Following that we have discussed the proposed hybrid loss function, which is the composite form of boundary loss, binary cross-entropy loss, dice loss, and squared hinge loss.

\subsection{Attention module architecture}
\begin{figure}
    \centering
    \includegraphics[scale=0.25]{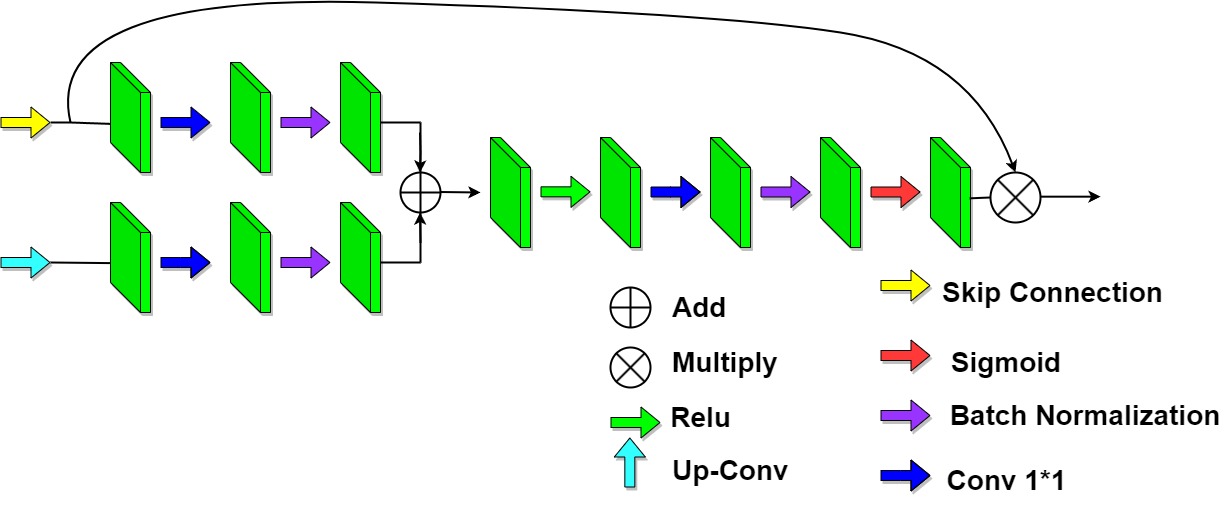}
    \caption{ Architecture of Attention mechanism}
    \label{attn}
\end{figure}

In Figure \ref{attn}, we have shown the attention mechanism's flow in the context of U-Net \cite{ronneberger2015u}. In contrast to existing network architectures, the attention module is utilized at the skip connections to focus on more relevant areas than the whole information. Such an attention method highlights important feature maps that determine the appropriate region or class and ignores unimportant ones. By using this method, the network is given the power to concentrate on portions of the image that are pertinent to a certain task.

The graphic representation (Fig \ref{attn}) demonstrates the flow of the attention mechanism as follows, our attention mechanism receives input from two separate locations, namely the network architecture's skip-connection and up-sampling component of the decoder. Both the convolution operation and the normalization are being applied to the coding characteristics from the two inputs. Then, to incorporate more feature information, the two parallel feature maps are combined at the pixel level. The single merged feature map is subjected to the following procedures, including Relu, $1 \times 1$ convolution, batch-normalization, and sigmoid function for activation. Finally, the output of this attention mechanism is calculated by multiplying it by the skip-connection's original feature map.

\subsection{Bi-category Hybrid Loss: A Composite Loss Function}
As COVID-19's tissue region tends to produce continuous zones that may be distinguished in a single CT slice. In order to prevent segmented regions from becoming fragmented and to ensure that they maintain the tissue's natural connectedness, COVID-CT-H-UNet uses the Bi-category Hybrid Loss function. This is crucial for things like lung tissue in CT images, where continuous swaths rather than discrete patches are anticipated to be seen. The Bi-category Hybrid loss consists of four loss functions of two categories. The weighted Binary Cross Entropy Loss  \cite{xie2015holistically} and the Square Hinge Loss  \cite{lee2013study} are two alternate loss functions that capture the pixel-wise similarities between predicted and ground truth images. On the other hand, Dice Loss \cite{sudre2017generalised} and Boundary loss  \cite{kervadec2019boundary} help in accurate boundary localization. More precisely, the Dice loss measures the overlap of the segmentation region between predicted and ground truth images. We have used a linear combination of these two categories of losses. The Formula for Bi-category Hybrid Loss(Bi-H loss) is given by:
$$
\text{Bi-H loss} = \alpha ( \operatorname{Weighted BC Loss + Dice Loss}) + \beta \operatorname{(Square Hinge Loss + Boundary Loss)}
$$
Empirically we have seen that Bi-category Hybrid Loss has the best segmentation impact on the COVID-CT-H-UNet network, where $\alpha$ and $\beta$ are hyper-parameters. However, the model's training process determines the optimal value for the coefficients $\alpha$ and $\beta$. Here $\alpha + \beta = 1$.

\section{Experimental result}
\begin{figure}
    \centering
    \includegraphics[width = \textwidth]{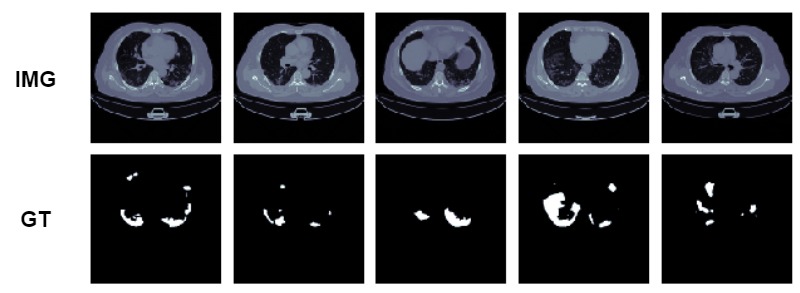}
    \caption{Few sample images and ground truths of the COVID-19 CT Segmentation dataset.}
    \label{data}
\end{figure}

There is a single COVID-19 CT segmentation dataset \cite{dataset} available due to the rarity of the dataset. Only $20$ CT scans make up this dataset for segmentation. Two radiologists recognized the left and right lungs, as well as any infections, and a seasoned radiologist confirmed their findings. CT scans are utilized for model testing by $20\%$ of CT scans while training takes up the other $80\%$. The affected region is shown in the ground truth as having intricate texture forms and characteristics, as seen in Figure \ref{data}. As a result, precisely segmenting the affected region from CT scans is quite challenging. 

\subsection{Experimental setup}
We have trained the Proposed COVID-CT-H-UNet from scratch in the Tensorflow framework with one 16GB NVIDIA Tesla P100 GPU. We have trained the network for around $100$ epochs with a batch size of $32$. Empirically we have seen that, the Adam optimizer performs better than SGD, hence we stuck to that and used a learning rate scheduler for the adaptive learning rate.

\subsection{Quantitative analysis} 
\subsubsection{Performance Metrics:}
In this paper, we have taken sensitivity, specificity, and dice coefficient as the evaluation metric for the segmentation. A detailed description of these metrics is given below.

The dice coefficient, which is frequently employed in medical image segmentation, is the ratio of the coincidence area of the double prediction map and ground truth to the total of the prediction map and ground truth.

\[ Dice =\frac
{2|A\cup B|}{|A + B|}
\]

where $A$ is the affected area in the ground truth.
$B$ is the predicted affected area in the prediction map.

The proportion of correctly labelled COVID-19 pixels to the total COVID-19 pixels is known as sensitivity.

\[
Sensitvity =\frac{T P}{T P + F N}
\]

where $TP$ in this instance stands for True Positive in which a genuine positive pixel, is accurately identified as COVID-19. False negative ($FN$) refers to the number of pixels that were mistakenly labelled as non-COVID-19.

The ratio of accurately designated non-COVID-19 pixels to the total number of real non-COVID-19 pixels is known as specificity.

\[
Specif icity =\frac{T N}{T N + F P}
\]

Here, $FP$ stands for false positive, which refers to the pixel that was mistakenly labeled as COVID-19, and TN stands for true negative, which refers to the pixels that were correctly recognized as non-COVID-19.

\subsubsection{Comparison with other segmentation networks:}
The traditional segmentation networks like U-Net \cite{ronneberger2015u}, U-Net++ \cite{zhou2018unet++}, U-Net+ResNet (i.e. UNet with ResNet backbone), TV-UNet \cite{saeedizadeh2021covid} and SCTV-UNet \cite{liu2023sctv} which have recently shown promising results on the COVID-19 CT segmentation dataset were chosen for comparison in this study. The comparison results are shown in Table \ref{acc_table}.

\begin{table}[h]

\centering
\caption{Comparison of indicators of different models in the COVID-19 CT segmentation dataset, bold indicates the best effect. Note that, except for SCTV-UNet \cite{liu2023sctv}, we have implemented all the models ourselves.}
\begin{tabular}{l|c|c|c}
\hline
\textbf{Model} & \textbf{Dice  }          & \textbf{Sensitivity}     & \textbf{Specificity }    \\ \hline
UNet \cite{ronneberger2015u}                        & 0.6048          & 0.7231          & 0.9996          \\ 
UNet++  \cite{zhou2018unet++}                     & 0.7712          & 0.6366          & 0.9995          \\ 
U-Net+ResNet                & 0.7856          & 0.7199          & 0.9997          \\ 
TV-UNet  \cite{saeedizadeh2021covid}                     & 0.7227          & 0.7032          & 0.9996          \\ 
SCTV-UNet  \cite{liu2023sctv}                     & 0.7989         & \textbf{0.8080}         & 0.9663       \\
\textbf{Proposed}           & \textbf{0.8947} & 0.7377 & \textbf{0.9997} \\ \hline
\end{tabular}%
\label{acc_table}
\end{table}

Compared to vanilla U-Net \cite{ronneberger2015u}, its improved versions like U-Net++, and U-Net+ResNet, perform better. On the other hand, the most recent networks like TV-UNet  \cite{saeedizadeh2021covid} and SCTV-UNet \cite{liu2023sctv} perform significantly superior to basic U-Net-based models. However, our proposed model outperforms all of them by a significant margin in Dice and Specificity metrics. Whereas SCTV-UNet \cite{liu2023sctv} has the best performance in the Sensitivity metric. One possible reason for such bad performance would be the higher number of false negative cases. Since the affected areas are small for a few images, the effect of BCE loss(as it is a part of the proposed Bi-H loss) can result in wrongly classified pixels leading to False negatives. In the future, we would like to revisit this problem by penalizing the negative effects of BCE loss.

\subsubsection{Impact of Bi-category Hybrid Loss function:} 
One of the key contributions of this paper is the Bi-category Hybrid Loss. According to Section 3.3, this paper's Bi-category Hybrid Loss solution attempts to address the issue of boundary blurring and poor foreground/background contrast in the prediction picture of COVID-19 CT segmentation using a U-shaped network. We have used other commonly used loss functions like Binary Cross Entropy(BCE) loss  \cite{xie2015holistically}, Dice Loss \cite{sudre2017generalised}, Boundary Loss \cite{kervadec2019boundary} etc. individually with our attention-UNet based segmentation model and compared with our proposed Bi-category Hybrid loss.  The performance results of dice, sensitivity, and specificity metrics are displayed in Table \ref{acc-table2},

\begin{table}[]
\centering
\caption{Model performance for different Loss Functions employed, bold indicates the best effect.}
\begin{tabular}{l|c|c|c}
\hline
\textbf{Loss} & \textbf{Dice   }         & \textbf{Sensitivity   }  & \textbf{Specificity }    \\ \hline
BCE                        & 0.8463          & 0.7267          & 0.9997          \\ \hline
DiceLoss + BoundaryLoss    & 0.8757          & 0.7186          & 0.9997          \\ \hline
BCE + DiceLoss             & 0.8820          & 0.7309          & 0.9996          \\ \hline
\textbf{Bi-H loss}          & \textbf{0.8947} & \textbf{0.7377} & \textbf{0.9997} \\ \hline
\end{tabular}

\label{acc-table2}
\end{table}

\subsection{Qualitative analysis}
As mentioned earlier, the COVID-19 CT segmentation dataset  \cite{dataset} dataset is used to train U-Net \cite{ronneberger2015u}, U-Net + Resnet, and COVID-CT-H-UNet. The following Figure \ref{network_comp} displays the test set's segmentation results.  It is evident that as the network gets better, the segmentation effect likewise gets better. However, there is an issue with incorrect detection of nearby lesion regions. That is why an attention method was introduced by COVID-CT-H-UNet. With the use of Loss composite loss function, the sick area could be emphasized while the unaffected area is hindered. As a result, the segmented infection region is more accurate.
\begin{figure}
    \centering
    \includegraphics[width = \textwidth]{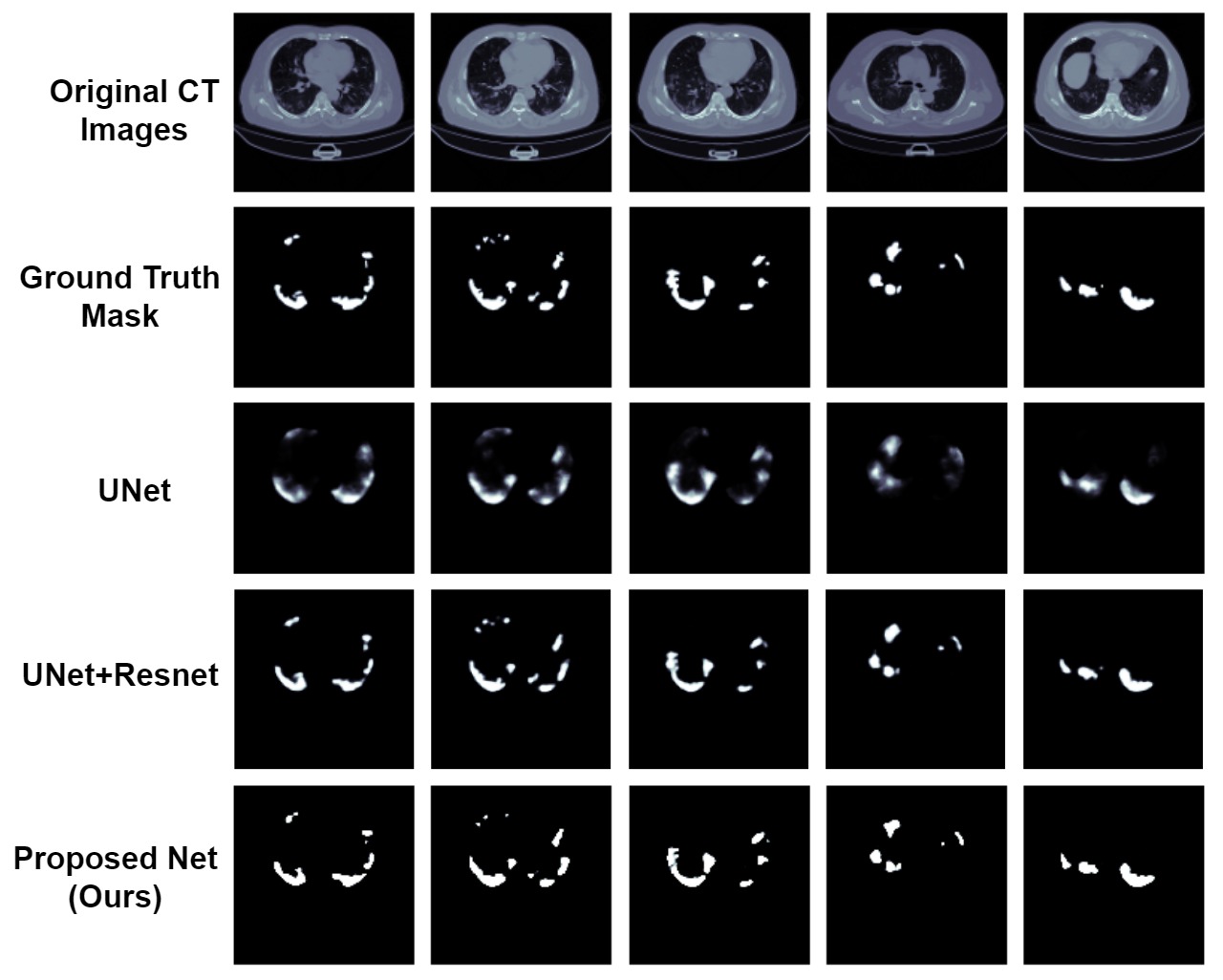}
    \caption{Qualitative comparisons of our models with the UNet and UNet+ResNet model}
    \label{network_comp}
\end{figure}

An attention mechanism is added to the skip-connections by COVID-CT-H-UNet. From Section 3.2, it is clear that the attention mechanisms enhance segmentation accuracy by allowing the model to assign varying levels of importance to different regions, effectively capturing intricate structures and subtle patterns that might be indicative of diseases or abnormalities. By adaptively highlighting significant information while suppressing noise, attention mechanisms contribute to more precise and robust segmentation outputs. The attention mechanism added to the skip-connections actually helps to gather more important semantic information and can emphasize the characteristics of the lesion region, and other areas. As a result, the issue wherein positive samples are mistaken for negative samples when there are contiguous lesion regions, which is a drawback of TV-UNet \cite{saeedizadeh2021covid} pointed out by SCTV-UNet  \cite{liu2023sctv}, is resolved. COVID-CT-H-UNet hence has a greater segmentation effect and accuracy.

The impact of COVID-CT-H-UNet's loss function design is seen in Fig. \ref{network_comp}.
The best segmentation outcomes come from the model's performance using a weighted combination of BCELoss, DiceLoss, Square Hinge Loss, and Boundary Loss. It is apparent that the issues of a hazy border and a lack of contrast between the lesion region and backdrop in TV-UNet \cite{saeedizadeh2021covid} prediction pictures have been resolved after the introduction of Loss.

\section{Conclusions}
This paper proposes a novel image segmentation network COVID-CT-H-UNet, which utilizes an attention network at the decoder part and a newly proposed Bi-category Hybrid loss for COVID-19 CT images. The attention block at the decoder helps to decode the relevant features from the previous steps along with the skip connection information to produce more accurate segmentation results. The newly proposed Bi-category Hybrid loss utilizes multiple loss functions effectively to increase the pixel similarity and the accurate boundary localization in the predicted segmented images. The proposed network is compared with other traditional networks in the field and has performed significantly better in terms of Dice score, Sensitivity, and Specificity. 
In the future, we'll implement this model in other fields and utilize dilated convolutions to make the model smaller in size. In future research, we'll also incorporate our Bi-category Hybrid loss with other segmentation and classification models to establish the superiority of the loss functions.

\begin{credits}
\subsubsection{\ackname} This research was carried out at the Centre for Microprocessor Application for Training Education and Research Lab within the Computer Science and Engineering Department of Jadavpur University and received partial support from a project funded by SERB, Govt. of India,(No: EEQ/2018/000963) and DST, GOI through the INSPIRE Fellowship program (IF170641).

\subsubsection{\discintname}
The authors have no conflict of interest to disclose.
\end{credits}

 \bibliographystyle{plain} 
 \bibliography{ref.bib}

\begin{thebibliography}{10}

\bibitem{dataset}
Covid-19 ct segmentation dataset. https://medicalsegmentation.com/covid19/.
\newblock 2020.

\bibitem{berman2018lovasz}
Maxim Berman, Amal~Rannen Triki, and Matthew~B Blaschko.
\newblock The lov{\'a}sz-softmax loss: A tractable surrogate for the optimization of the intersection-over-union measure in neural networks.
\newblock In {\em Proceedings of the IEEE conference on computer vision and pattern recognition}, pages 4413--4421, 2018.

\bibitem{chen2021transunet}
Jieneng Chen, Yongyi Lu, Qihang Yu, Xiangde Luo, Ehsan Adeli, Yan Wang, Le~Lu, Alan~L Yuille, and Yuyin Zhou.
\newblock Transunet: Transformers make strong encoders for medical image segmentation.
\newblock {\em arXiv preprint arXiv:2102.04306}, 2021.

\bibitem{chen2018encoder}
Liang-Chieh Chen, Yukun Zhu, George Papandreou, Florian Schroff, and Hartwig Adam.
\newblock Encoder-decoder with atrous separable convolution for semantic image segmentation.
\newblock In {\em Proceedings of the European conference on computer vision (ECCV)}, pages 801--818, 2018.

\bibitem{fan2020inf}
Deng-Ping Fan, Tao Zhou, Ge-Peng Ji, Yi~Zhou, Geng Chen, Huazhu Fu, Jianbing Shen, and Ling Shao.
\newblock Inf-net: Automatic covid-19 lung infection segmentation from ct images.
\newblock {\em IEEE transactions on medical imaging}, 39(8):2626--2637, 2020.

\bibitem{hu2018squeeze}
Jie Hu, Li~Shen, and Gang Sun.
\newblock Squeeze-and-excitation networks.
\newblock In {\em Proceedings of the IEEE conference on computer vision and pattern recognition}, pages 7132--7141, 2018.

\bibitem{ibtehaz2020multiresunet}
Nabil Ibtehaz and M~Sohel Rahman.
\newblock Multiresunet: Rethinking the u-net architecture for multimodal biomedical image segmentation.
\newblock {\em Neural networks}, 121:74--87, 2020.

\bibitem{kervadec2019boundary}
Hoel Kervadec, Jihene Bouchtiba, Christian Desrosiers, Eric Granger, Jose Dolz, and Ismail~Ben Ayed.
\newblock Boundary loss for highly unbalanced segmentation.
\newblock In {\em International conference on medical imaging with deep learning}, pages 285--296. PMLR, 2019.

\bibitem{lee2013study}
Ching-Pei Lee and Chih-Jen Lin.
\newblock A study on l2-loss (squared hinge-loss) multiclass svm.
\newblock {\em Neural computation}, 25(5):1302--1323, 2013.

\bibitem{li2018h}
Xiaomeng Li, Hao Chen, Xiaojuan Qi, Qi~Dou, Chi-Wing Fu, and Pheng-Ann Heng.
\newblock H-denseunet: hybrid densely connected unet for liver and tumor segmentation from ct volumes.
\newblock {\em IEEE transactions on medical imaging}, 37(12):2663--2674, 2018.

\bibitem{liu2023sctv}
Xiangbin Liu, Ying Liu, Weina Fu, and Shuai Liu.
\newblock Sctv-unet: a covid-19 ct segmentation network based on attention mechanism.
\newblock {\em Soft Computing}, pages 1--11, 2023.

\bibitem{long2015fully}
Jonathan Long, Evan Shelhamer, and Trevor Darrell.
\newblock Fully convolutional networks for semantic segmentation.
\newblock In {\em Proceedings of the IEEE conference on computer vision and pattern recognition}, pages 3431--3440, 2015.

\bibitem{lu2022multiscale}
Q~Lu, Z~Bai, S~Fan, X~Zhou, and Z~Xu.
\newblock Multiscale codec network based ct image segmentation for human lung disease derived of covid-19.
\newblock {\em Journal of Image and Graphics}, pages 827--837, 2022.

\bibitem{miller20202019}
Meg Miller.
\newblock 2019 novel coronavirus covid-19 (2019-ncov) data repository: Johns hopkins university center for systems science and engineering.
\newblock {\em Bulletin-Association of Canadian Map Libraries and Archives (ACMLA)}, (164):47--51, 2020.

\bibitem{paules2020coronavirus}
Catharine~I Paules, Hilary~D Marston, and Anthony~S Fauci.
\newblock Coronavirus infections—more than just the common cold.
\newblock {\em Jama}, 323(8):707--708, 2020.

\bibitem{ronneberger2015u}
Olaf Ronneberger, Philipp Fischer, and Thomas Brox.
\newblock U-net: Convolutional networks for biomedical image segmentation.
\newblock In {\em Medical Image Computing and Computer-Assisted Intervention--MICCAI 2015: 18th International Conference, Munich, Germany, October 5-9, 2015, Proceedings, Part III 18}, pages 234--241. Springer, 2015.

\bibitem{saeedizadeh2021covid}
Narges Saeedizadeh, Shervin Minaee, Rahele Kafieh, Shakib Yazdani, and Milan Sonka.
\newblock Covid tv-unet: Segmenting covid-19 chest ct images using connectivity imposed unet.
\newblock {\em Computer methods and programs in biomedicine update}, 1:100007, 2021.

\bibitem{sluimer2006computer}
Ingrid Sluimer, Arnold Schilham, Mathias Prokop, and Bram Van~Ginneken.
\newblock Computer analysis of computed tomography scans of the lung: a survey.
\newblock {\em IEEE transactions on medical imaging}, 25(4):385--405, 2006.

\bibitem{sudre2017generalised}
Carole~H Sudre, Wenqi Li, Tom Vercauteren, Sebastien Ourselin, and M~Jorge~Cardoso.
\newblock Generalised dice overlap as a deep learning loss function for highly unbalanced segmentations.
\newblock In {\em Deep Learning in Medical Image Analysis and Multimodal Learning for Clinical Decision Support: Third International Workshop, DLMIA 2017, and 7th International Workshop, ML-CDS 2017, Held in Conjunction with MICCAI 2017, Qu{\'e}bec City, QC, Canada, September 14, Proceedings 3}, pages 240--248. Springer, 2017.

\bibitem{wang2019learning}
Qiang Wang, Bei Li, Tong Xiao, Jingbo Zhu, Changliang Li, Derek~F Wong, and Lidia~S Chao.
\newblock Learning deep transformer models for machine translation.
\newblock {\em arXiv preprint arXiv:1906.01787}, 2019.

\bibitem{wang2021covid}
Shui-Hua Wang, Vishnu~Varthanan Govindaraj, Juan~Manuel G{\'o}rriz, Xin Zhang, and Yu-Dong Zhang.
\newblock Covid-19 classification by fgcnet with deep feature fusion from graph convolutional network and convolutional neural network.
\newblock {\em Information Fusion}, 67:208--229, 2021.

\bibitem{woo2018cbam}
Sanghyun Woo, Jongchan Park, Joon-Young Lee, and In~So Kweon.
\newblock Cbam: Convolutional block attention module.
\newblock In {\em Proceedings of the European conference on computer vision (ECCV)}, pages 3--19, 2018.

\bibitem{xie2015holistically}
Saining Xie and Zhuowen Tu.
\newblock Holistically-nested edge detection.
\newblock In {\em Proceedings of the IEEE international conference on computer vision}, pages 1395--1403, 2015.

\bibitem{zhou2018unet++}
Zongwei Zhou, Md~Mahfuzur Rahman~Siddiquee, Nima Tajbakhsh, and Jianming Liang.
\newblock Unet++: A nested u-net architecture for medical image segmentation.
\newblock In {\em Deep Learning in Medical Image Analysis and Multimodal Learning for Clinical Decision Support: 4th International Workshop, DLMIA 2018, and 8th International Workshop, ML-CDS 2018, Held in Conjunction with MICCAI 2018, Granada, Spain, September 20, 2018, Proceedings 4}, pages 3--11. Springer, 2018.

\end{thebibliography}

\end{document}